# Strong Coupling beyond the High-Q Limit and Linewidth Narrowing in a Multi-Exciton Planar Microcavity.

E. A. Cerda-Mendez[1,*], Y. G. Rubo[2], K. Biermann[3], A. Camacho-Guardian[4], A. S. Kuznetsov[3] and P. V. Santos[3].

[1]Instituto de Investigación en Comunicación Óptica, Universidad Autónoma de San Luis Potosí, San Luis Potosí, México
[2]Instituto de Energías Renovables, Universidad Nacional Autónoma de México, Temixco, Morelos, México.
[3]Paul-Drude-Institut für Festkörperelektronik im Forschungsverbund Berlin e.V., Berlin, Germany
[4]Instituto de Física, Universidad Nacional Autónoma de México, Ciudad de México, México.

**ABSTRACT**.
We systematically study the linewidths of multilevel exciton-polariton modes as a function of the detuning in a planar hybrid microcavity (MC) with low quality factor (Q~300) operating in the linear optical response regime. Using optical reflectivity, we observe that, counterintuitively, the linewidths of the polariton modes undergo a pronounced spectral narrowing as detuning is reduced. By benchmarking the experimental results against three commonly used constant-loss theoretical descriptions, we find that this behavior is only partially reproduced, highlighting limitations of conventional strong-coupling models when applied to non-standard MC architectures. Our results suggest that frequency-dependent self-energy effects or correlated dissipation mechanisms, typically neglected in simplified treatments, may play an important role. These observations indicate that strong light–matter coupling with narrow polariton lines can be achieved in hybrid low-$Q$ MCs through appropriate loss engineering, providing an alternative design pathway complementary to conventional high-$Q$ approaches.

## I. INTRODUCTION.

The study of strong coupling (SC) of light and matter has emerged as a novel venue in a plethora of platforms since its discovery in planar semiconductor microcavities (MC) [1]. The observation of macroscopic polariton quantum phases [2–4] and related phenomena such as superfluidity, quantum vorticity and soliton formation, among others [5,6] have motivated multiple proposals and demonstrations of novel devices for information processing [7,8], physical systems analogues [9,10] study of coherent phenomena [11] and efficient emission of coherent light [12], for example. This huge potential in a solid-state platform motivates the study of SC in semiconducting systems beyond III-V materials, which require cryogenic temperatures for operation and demanding technological facilities for their fabrication. These include 2D [13,14], molecular [15], perovskite [16] and even biological materials [17]. Consequently, novel technological platforms to fabricate MCs are continuously being developed. Also, several studies on multilevel polariton systems have been done demonstrating interesting phenomena such as SC with heavy- and light-hole excitons [18], Faraday rotation [19], SC and Bose-Einstein condensation of polaritons in (Al,Ga)As-based MCs at room temperature [20,21] and slow-light physics [22].

SC in MCs was originally studied in a monolithic (Al,Ga)As multilayer structure composed of two opposing distributed Bragg reflectors (DBR), separated by a spacer layer containing GaAs quantum wells (QW) [1]. This architecture, inherited from vertical cavity surface emitting lasers (VCSELs), confines the optical field between the DBRs while the QWs host excitons (bound electron–hole pairs). The repeated coherent absorption and re-emission of photons by these excitons within the spacer layer produces an interaction that mixes the excitonic and photonic modes, giving rise to new quasiparticles known as exciton–polaritons. The ability of the MC to store electromagnetic energy is characterized by its quality factor $Q = E_{ph}/\delta E_{ph}$, where $E_{ph}$ is the energy of the optical resonance and $\delta E_{ph}$ is its linewidth, defined as the full width at half maximum (FWHM) of the resonance. While VCSELs are designed to maximize stimulated emission and achieve population inversion, polaritonic MCs rely on

*Contact author: edgar.cerda@uaslp.mx

photon–exciton recycling to sustain SC. Accordingly, standard GaAs-based VCSELs have $Q \approx 10^3$–$10^4$ and III–V polariton MCs typically operate at cryogenic temperatures with $Q \gtrsim 10^3$ to reach the SC regime. In general, the polariton condensation threshold is lower than the regular lasing threshold in MCs [23,24], leading to a prevailing design principle for polaritonic devices: "the higher the $Q$, the better". It is thus usually assumed that a long-living photonic mode is necessary to achieve the SC regime in planar MC. This belief is mainly based on the consideration of a simple two coupled-oscillators Hamiltonian for photons and excitons, that demonstrates that SC occurs when the Rabi frequency Ω is larger than the dissipation rate of the cavity $\gamma_{cav}$. However, this model does not consider factors such as the inhomogeneous broadening of the exciton resonance and possible motional narrowing of polariton lines [25–27]. For example, for large excitonic inhomogeneous broadening it might be better to have a comparably broad photonic line, so that photons are well coupled to all the spread of exciton levels. Similarly, a narrow photonic mode could be undesirable in the case when several types of excitons with different frequencies are present.

Linewidth narrowing of polaritons is a compelling phenomenon often attributed to motional narrowing [25–30], where SC between a high-$Q$ photonic mode and an inhomogeneously broadened excitonic resonance give rise to delocalized polariton states that average out the inhomogeneity. In this work we experimentally demonstrate a distinct mechanism: the narrowing by nearly a factor of four across the detuning range, of a homogeneously broadened photonic single mode in a planar MC with low $Q \approx 200 - 300$. As the broad photon mode approaches resonance with the narrow heavy- and light-hole QW exciton resonances, three well-resolved polariton branches form. Importantly, the total linewidth of the three branches decreases with detuning, challenging the trace conservation typically expected from a conventional interaction Hamiltonian, where total linewidth is preserved. Our results demonstrate that SC can occur and even produce linewidth narrowing in low-$Q$ cavities when coupling multiple excitonic resonances. Although exciton polaritons in GaAs microcavities are a well-established platform, the linewidth behavior reported here emerges in a regime combining low cavity quality factor and simultaneous coupling to heavy- and light-hole excitons that has not been systematically explored.

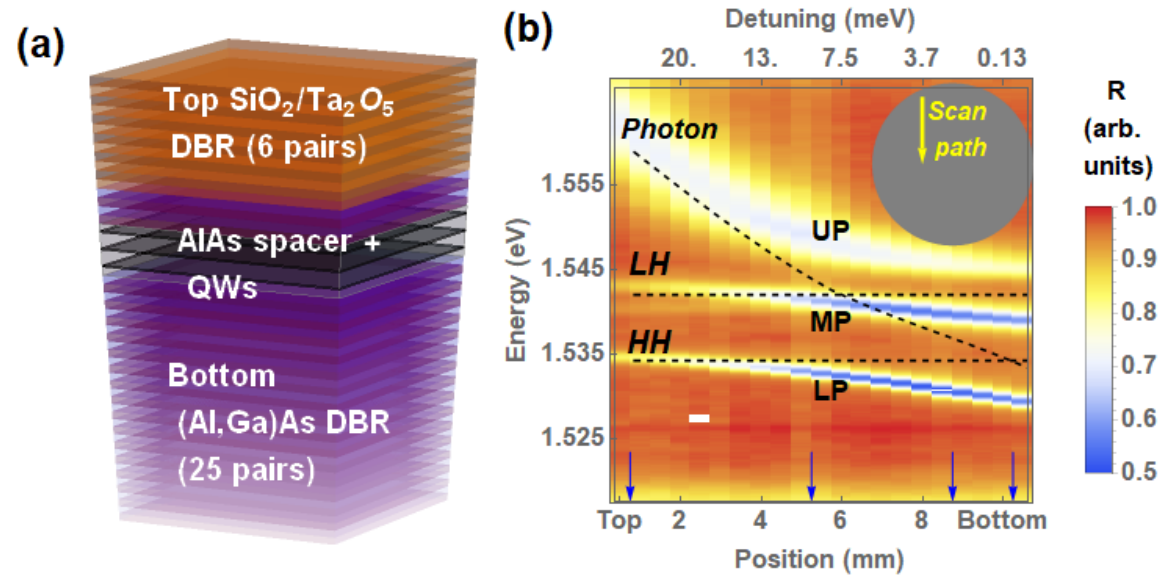

FIG 1. a) Schematic representation of the hybrid dielectric-semiconductor MC. b) Experimental evidence of linewidth narrowing in low-Q cavity polaritons. The plot shows a spatially resolved reflectance dataset comprising twenty individual spectra acquired across a continuous detuning gradient between the photon and heavy-hole (HH) exciton resonance which is a function of the position on the sample wafer along the scan path (inset). The complete spatial scan from which these data are derived is shown in Fig. A1 of Appendix A. The characteristic avoided-crossing behavior due to polariton formation with both heavy- and light-hole (LH) excitons is clearly visible. The black dashed lines show the energies of the HH and LH excitons and the pure photonic mode. The photon energies are calculated from the transfer matrix simulations discussed in the text. Inset: the grey circle represents the 2-inch wafer, and the arrow is the path followed to make the measurements. The blue arrows show the positions where the spectra in Fig. 2 were taken.

We show as well that this behavior can be only partially reproduced by commonly used constant-decay rates approaches such as interaction Hamiltonian diagonalization, transfer matrix method (TMM) simulation or non-perturbative Green's function approach evaluating the spectral response function $\mathcal{A}(\omega)$. While motional narrowing may play a role, our findings suggest that SC to excitonic states suppresses radiative broadening of the photon mode itself, enhancing coherence even in lossy cavities. This inversion of the conventional paradigm—where higher $Q$ is presumed necessary for coherent polariton operation—has not been previously demonstrated, to the best of our knowledge. Although SC has been shown in low-$Q$ plasmonic systems [33–37], these rely on ultra-small mode volumes. Our results thus call for a reassessment of cavity $Q$-based design principles in polaritonics and suggest a novel regime where coherence is enhanced via coupling rather than cavity finesse, opening new avenues for unconventional polariton platforms and applications.

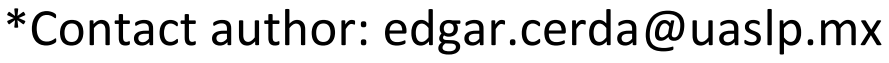

## II. EXPERIMENTAL ASPECTS

The hybrid MC sample is composed of an (Al,Ga)As-based semiconductor half-MC (bottom DBR and spacer layer containing three groups of three 15 nm GaAs QWs) and a top $SiO_2/Ta_2O_5$ dielectric DBR (see Fig. 1(a) and Appendix B). The number of QWs substantially enhances the light–matter coupling strength and their relatively large thickness results in a non-negligible contribution from both heavy- and light-hole excitons, in contrast to more commonly studied two-level polariton systems dominated solely by the heavy-hole transition. The hybrid MC architecture departs from conventional monolithic III–V DBR structures and is particularly relevant for emerging polaritonic platforms where cavity losses and mode profiles differ significantly from standard designs. The bottom half-MC was grown by standard molecular beam epitaxy (MBE). In samples routinely synthesized in our system on 2-in diameter GaAs wafers, we observe a spatial gradient in the layer thicknesses of about 2 % from the center to the border of the wafer due to the geometry of the MBE reactor. By careful growth engineering, it is possible to fine tune the gradient to get thicker layers at the center, which offer a lower photonic energy resonance, and thinner ones at the borders, which lead to a higher photonic energy resonance. The thickness gradient, though, does not change the QW thickness significantly, so the exciton energy remains nearly constant across the sample. This is a powerful tool as a broad range of detuning values $\Delta E = E_{ph} - E_{exc}$ between the photonic ($E_{ph}$) and excitonic ($E_{exc}$) resonances becomes accessible, allowing variation of the excitonic and photonic components in the polariton branches by changing the location of the measurement on the sample. Since in our system we observe coupling of light with both heavy- and light-hole exciton levels, we define the detuning $\Delta E$ with respect to the lowest-energy heavy-hole exciton level.

The top dielectric DBR is deposited on the semiconductor half-MC by radio frequency assisted sputtering (rf-sputtering) of $SiO_2$ and $Ta_2O_5$. Due to the higher refractive index contrast, nominally 6 pairs are sufficient to achieve high reflectivity. While rf-sputtering is a standard technique to grow high quality thin layers, it is very challenging to fabricate high-quality DBRs with it, even with a small number of layers as in our case. Despite a careful previous calibration of the growth rate, the fabricated layers had a smaller optical thickness $nd$ than planned, due to either smaller thickness $d$ or refractive index $n$, or a combination of both. As a result, the $Q$ factor of the MC terminated with the dielectric DBR is much lower than expected (200-300), well below typical thresholds for SC in planar cavities. Nevertheless, as discussed below, pronounced Rabi splitting and linewidth redistribution are observed, indicating that the coupling exceeds the dissipative rate of the cavity mode.

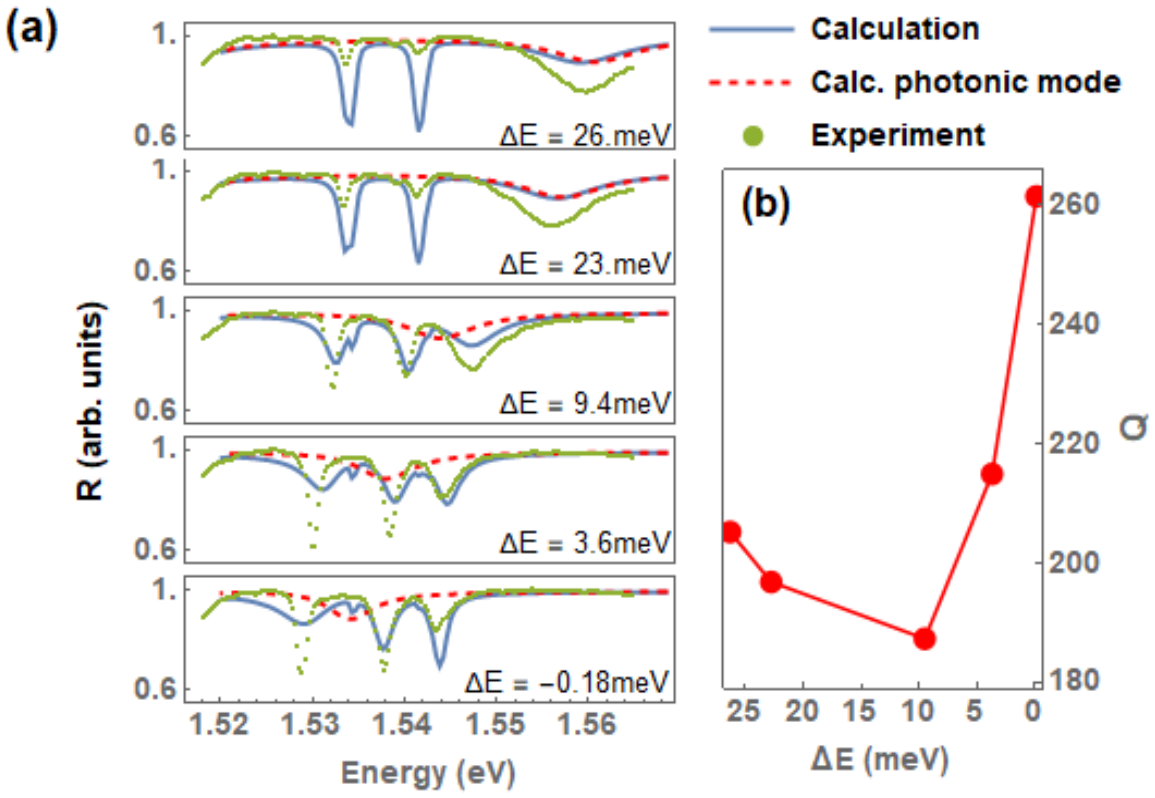

FIG 2. (a) Transfer matrix method (TMM) calculation (blue lines) of the reflectivity spectra (green dots) measured at the positions marked by the blue arrows in Fig. 1(b). The calculated value of detuning $\Delta E$ is shown in each panel. The black dashed line is the pure photonic mode extracted from the TMM calculation by removing the exciton oscillators absorption. (b) $Q$ factor of the calculated photonic mode for the different detuning values $\Delta E$ in panel (a).

The reflectivity measurements were carried out by focusing a tungsten filament lamp into a 500 $\mu m$ spot with a 10x microscope objective. The reflected light was collected and collimated with the same objective and spatially filtered in the reciprocal image plane with a pinhole to select the beams close to normal ($k_{||} \approx 0$), thus avoiding artificial broadening due to the cavity dispersion. A spatial scan was made by displacing a microscope objective stage in steps of 0.5 mm along a 1 cm path along the wafer in a cryostat at 15 K (inset Fig. 1 (b)). The reflectivity spatial scan comprised of twenty individual spectra (Fig. 1(b)) reveals that, as the low-$Q$ photonic resonance energy redshifts when moving towards the center of the sample, it gets narrower and eventually shows the characteristic avoided crossing behavior with the excitonic light- and heavy-hole transitions, signaling the presence of SC and polariton formation. Importantly, the linewidths of the middle and lower branches do not increase significantly. Note that for smaller detuning values (larger $y$-position values in the plot), the energy of both lower polariton branches decreases, showing coupling of the photon with both light- and heavy-hole QW excitons. Also, the red shift of the excitonic resonances cannot be ascribed solely to the QWs being thicker at the center. A change of thickness of 2% (0.3 nm) corresponds to less than 1 meV, which is of the order of the linewidth and thus negligible at this scale. Although the effect is observed within a single MC, it is revealed through a continuous and reproducible scan over detuning, providing a statistically meaningful and systematic dataset rather than a single-point observation. The full reflectivity map of the wafer area from where Fig. 1 (b) was taken is shown in Fig. A1 of Appendix A.

Finally, we note that the linewidth narrowing is observed under linear reflectance conditions, with excitation densities many orders of magnitude below those required for polariton condensation or stimulated emission in comparable GaAs-based microcavities. This ensures operation in the linear optical response regime, where interaction- or density-dependent mechanisms are negligible.

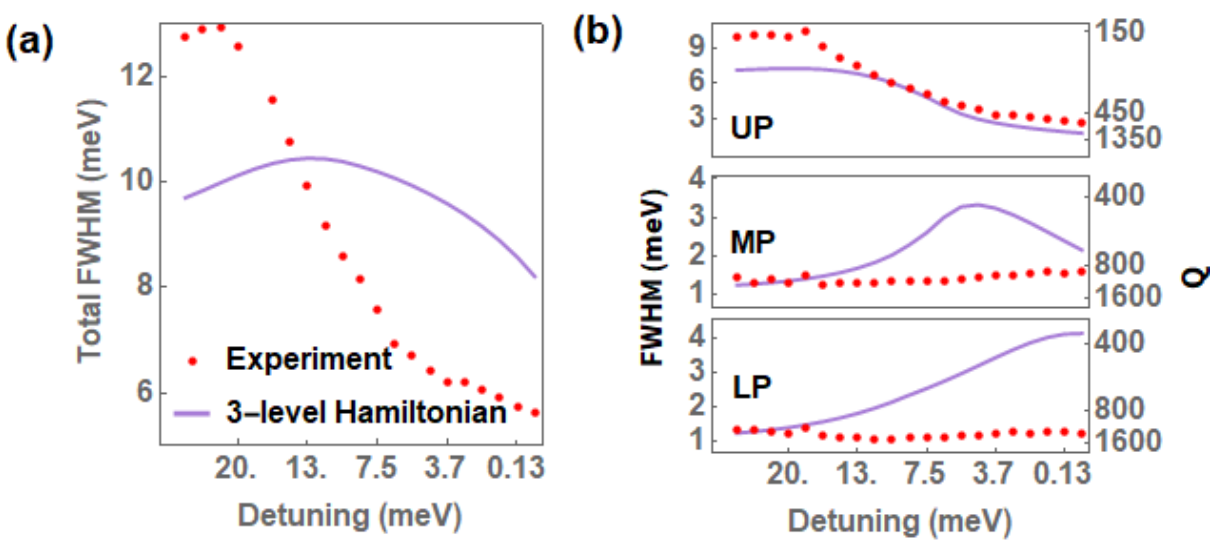

FIG 3. Comparison of a) the total and b) the individual upper-, middle-, and lower-polariton branches full-widths at half-maximum (FWHM). The experimental points are extracted from Fig. 1(b) (red dots) and the calculation (thick purple line) from direct diagonalization of the three-level Hamiltonian. In the panels of b), the right axes show the equivalent $Q$-factor.

## III. ANALYSIS OF RESULTS

We discuss now in detail the evolution of the position and linewidths of the polariton branches with detuning. For this, we take a few representative experimental spectra from Fig. 1(a) at the positions marked by the blue arrows, shown in green dots in Fig. 2(a). We then calculate the reflectivity coefficient of the MC at these positions using TMM. The physical parameters of the top dielectric DBR are extracted from reflectivity measurements over a broad-wavelength range (see Appendix B). To account for the inhomogeneous broadening of the excitons, we averaged spectra calculated with excitonic Lorentzian oscillators with infinitesimal linewidth distributed from the central energy by an amount equivalent to the measured excitonic linewidths of 1 meV and 1.5 meV for the heavy- and light-holes, respectively. This method provides a good semiclassical description of the motional narrowing in the case of single exciton resonances [26]. The results of the full TMM calculation are shown in blue continuous lines in Fig. 2(a). Note the dashed red lines as well, which show the corresponding pure optical resonances calculated by removing the imaginary part (absorption) of the exciton oscillators from the full TMM result. The calculated photonic modes show that the lineshape and energy of the photonic resonance is dependent on the detuning, which, as mentioned, varies with position. At the largest detuning, the upper branch is mostly photonic with $Q \simeq 200$. As the detuning decreases, the purely optical $Q$-factor first decreases slightly and then shows a modest increase to a value of $Q \simeq 280$. This variation is explained by the thickness gradient of the bottom semiconductor half-cavity layers: their thickening towards the center of the substrate wafer redshifts the bottom DBR stopband, realigning it with the broader stopband of the top dielectric DBR (which does not depend on the position). The concomitant variation of the thickness of the spacer layer shifts the resonance wavelength as well. We note that the only free parameter in the TMM calculation is the uniform change of thickness of the layers of the bottom semiconductor half-cavity. The energies and amplitudes of the excitonic Lorentzians are fixed for all the spectra. While we obtain a very good agreement between the experiment and the calculation regarding the position of the peaks (see Fig. C1(b)), the shapes, widths, and relative amplitudes of the lines are only partially reproduced.

The dependence of the FWHM on the detuning extracted from the experiment in Fig. 1(b) is shown in red dots in Figs. 3(a) (total FWHM of the branches) and 3(b) (individual branches). This dependence is non-trivial. For example, the strong and monotonical reduction of the total FWHM and of the upper branch with detuning shows no correlation to the dependence of the photonic $Q$-factor in Fig. 2(b). The narrowing of the total FWHM by 52% and of the upper branch by 67% thus cannot be accounted for by the improvement of the $Q$-factor of the photonic mode alone. Importantly, the linewidths of the lower and middle polariton branches remain almost constant for all detuning values. We note that the total FWHM is introduced as a diagnostic quantity to compactly highlight discrepancies between experiment and theory, rather than as a fundamental observable or a proposed device metric.

We study these experimental observations with two commonly used models: a simple 3-level Hamiltonian with complex energies to account for line broadening, and calculation of the spectral response function $\mathcal{A}(\omega) = -Im\,[G_{cc}(\omega)]$, where $G_{cc}(\omega)$ is the retarded photon Green's function at $k = 0$. $G_{cc}(\omega)$ incorporates the self-energy correction due to the coupling with heavy- and light-hole excitonic resonances (see Appendix C). The calculation of $\mathcal{A}(\omega)$ allows the computation of the full polaritonic response, naturally incorporating finite linewidths, energy-dependent responses, yielding realistic spectral lineshapes suitable for direct comparison with experiments. [27]

From the TMM calculation, we obtain numerical interpolation functions of the dependence of the photon energy $E_{ph}(y)$, and its linewidth $\gamma_{ph}(y)$ with position $y$ (see Fig. C2). This, along with the fixed, experimentally measured bare heavy- and light-hole excitons energies $(E_{hh}, E_{lh})$ and linewidths $(\gamma_{hh}, \gamma_{lh})$ are then fed into a simple three-level model for polariton branches, defined by the complex Hamiltonian

$$\mathcal{H}(y) = \begin{pmatrix} E_{ph}(y) - i\gamma_{ph}(y) & \Omega_{lh} & \Omega_{hh} \\ \Omega_{lh} & E_{lh} - i\gamma_{lh} & 0 \\ \Omega_{hh} & 0 & E_{hh} - i\gamma_{hh} \end{pmatrix} \quad (1)$$

By making $\gamma_{ph}, \gamma_{hh}, \gamma_{lh} = 0$, we can estimate the values of the interaction terms $\Omega_{hh} = 10\, meV$ and $\Omega_{lh} = 6.7\, meV$ which quantify the Rabi splitting of the photon and the respective exciton. (see Appendix C and Fig. C2).

A diagonalization of the interaction Hamiltonian (1) to account for the broadening is insufficient to understand the observed FWHM dependences on detuning in Fig. 3: it overestimates the total linewidth $\gamma_{tot} = \gamma_{ph} + \gamma_{hh} + \gamma_{lh}$ of the branches by as much as 37% (panel (a)). Moreover, it leads to qualitatively different behavior of $\gamma_{tot}$ as function of detuning, which is closer to that of $Q$ in Fig. 2(b). Note that if $Q$ were position independent, $\gamma_{tot}$ in this model would be constant (i.e. independent of detuning) because of the conservation of the trace of the Hamiltonian matrix. Also, it should remain at least as big as the free photon linewidth.

Both the diagonalization of the complex Hamiltonian and the Green's-function calculation rely on the same secular equation—the poles of the retarded photon propagator coincide exactly with the complex eigenvalues of the $3 \times 3$ matrix. Therefore, both approaches are isospectral and should yield identical mode energies and linewidths when the loss rates are constant. The difference between them lies mainly in the interpretation of linewidths and residues: while the Hamiltonian picture treats damping phenomenologically, the Green's-function framework allows extension to frequency-dependent self-energies $\Sigma(\omega)$, which become essential once strong dissipation or spectral overlap of heavy- and light-hole excitons is considered. Also, in the linear excitation regime relevant to our reflectance measurements, a conventional Lindblad master-equation description with Markovian loss channels is formally isospectral to the non-Hermitian Hamiltonian and Green's-function approaches discussed here. Consequently, all such constant-loss with uncorrelated and frequency-independent dissipation channels models face the same limitations in accounting for the observed linewidth redistribution and narrowing. The formal equivalence between the three formalisms ensures that the observed discrepancy—the narrowing of the cavity-like mode—must originate from physics not captured by a constant-loss model, rather than from the mathematical formulation itself. The results of the Green's-function calculation, including a full simulation of the experiments in Fig. 1(b), are summarized in Appendix C. As with the complex Hamiltonian (1), the data used are those extracted from the TMM calculation.

## IV. DISCUSSION

Our result thus demonstrates a substantial and counter-intuitive narrowing of the photon-like upper polariton branch as the detuning decreases, despite the broad linewidth of the uncoupled cavity mode (~10 meV), comparable to the Rabi coupling energy in these systems. While motional narrowing may also contribute to the narrowing of the LP branch in our system, the key observation is that SC to a distribution of excitonic states seems to lead to a suppression of radiative broadening of the photonic mode itself, effectively acting as a coherence-enhancing mechanism for a lossy cavity. To our knowledge, this inversion of the conventional expectation—that high-$Q$ cavities are required to achieve narrower and more coherent polariton modes—has so far not been reported before in either GaAs-based or other multilevel polariton platforms [18,21,31,32].

To further clarify this point, we compared the three-level Hamiltonian with a Dyson Green's-function formalism, which rigorously includes the excitonic self-energies responsible for the radiative coupling. Both approaches are formally equivalent in the linear regime, yielding the same complex poles and, hence, identical polariton energies and intrinsic linewidths when the damping constants are fixed. However, neither fully reproduces the experimentally observed narrowing of the cavity-like branch, indicating that additional physical mechanisms must be at play. In this sense, the Green's-function approach provides a natural route to include such effects systematically, while the Hamiltonian model highlights the conservation-of-trace constraint that prevents linewidth redistribution in the constant -loss approximation. Since all measurements are performed in the linear reflectance regime, exciton–exciton interactions and many-body correlation effects are negligible. The observed linewidth redistribution must therefore originate from linear light-matter SC physics and modified dissipation and/or interference mechanisms arising from hybridization of multiple excitonic resonances with the cavity mode and related interference effects. These may include frequency-dependent self-energies $\Sigma(\omega)$, cross-correlated dissipation between photon and exciton reservoirs, or scattering between exciton states at different wave vectors [27], which could lead to nontrivial redistribution of radiative lifetimes due to destructive interference or reduced coupling to leaky channels [39,40]. This interpretation aligns with the expectation that, beyond the constant-loss approximation, off-diagonal elements in the self-energy tensor could lead to effective radiative lifetime redistribution, consistent with the observed narrowing at near-zero detuning. A full simulation including detailed study of the scattering processes in our experiment is beyond the scope of the present report. We explicitly acknowledge that a complete microscopic explanation of the observed linewidth reduction is not yet available. Rather than constituting a shortcoming, this represents a central outcome of the present work: the experimental demonstration that commonly used constant-loss descriptions of strong coupling are insufficient to account for the observed linewidth redistribution. While it can be claimed that, in general, SC remains significant as long as the linewidth of the photon is smaller than the coupling constant, it is not expected nor trivial to see that the resultant polariton linewidths become as narrow as observed. This means that the assumption regarding having a high $Q$ is useful only as a first approximation for the design of devices sustaining SC. By using estimations derived from this model alone, a device could be designed to have a $Q$-factor much higher than necessary: for, example, in the case of the system presented here, the three-level Hamiltonian predicts a linewidth of the lower polariton branch more than 3 times larger than observed. From an applied perspective, these results are relevant for the design of polaritonic microcavities operating beyond the conventional high-$Q$ regime, particularly in hybrid or integrated platforms where cavity losses and fabrication constraints limit achievable quality factors.

## V. CONCLUSIONS

In conclusion, we presented experimental evidence for a nontrivial narrowing of the linewidths of polariton branches in a three-level system, where photons mix with heavy- and light-holes QW excitons in a low-$Q$ planar microcavity. This observation cannot be explained by the commonly used interaction Hamiltonian, which can lead to gross underestimations of the lifetime of the polaritonic modes.

The possibility of achieving SC and linewidth narrowing in intentionally low-Q cavities challenges the conventional expectation that long photonic lifetimes are required for hybridization. From an applied perspective, this regime enables polaritonic behavior in so in non-conventional platforms such as the one presented here or those based on 2D materials, organic semiconductors, molecular polaritons, etc., where large inhomogeneous broadening and hybrid MC architectures are common. [13–16]. Further theoretical refinements using frequency-dependent self-energies could bridge the gap between this phenomenology and the microscopic origin of dissipation in realistic semiconductor heterostructures. Beyond fundamental interest, understanding linewidth redistribution in low-Q hybrid cavities is relevant for the engineering of polaritonic devices where cavity quality cannot easily be increased. In the growing field of SC in non-conventional platforms, this work demonstrates that tools beyond the ones commonly used are crucial for the design and optimization of devices to fully exploit their potential.

## VI. APPENDIX A: COMPLETE DATA SET

In this Appendix, we present a summary of the analysis performed on the complete set of reflectivity measurements from which the spectra in Figs. 1 and 2 are extracted (see Fig. A1 of Appendix A).

The reflectivity spectra were measured over an area of $5.0 \times 10.0\, mm^2$, with measurement points separated by 500 $\mu m$. In total, 220 individual spectra were acquired across the sample. Each spectrum was analyzed by fitting a model consisting of three Lorentzian oscillators, corresponding to the upper (UP), middle (MP), and lower polariton (LP) branches. Figure A1 summarizes the results of this analysis. The dots indicate the extracted resonance energies of each polariton branch at the corresponding spatial position, while the vertical gray bars represent the FWHM. For a small number of spatial positions in the UP branch, no reliable fits could be obtained; these points are therefore omitted from the dataset. The red dots are the data subset discussed in the manuscript.

## VII. APPENDIX B: METHODS

In this Appendix we discuss the sample design and fabrication and characterization methods. DBRs are stackings of multiple repetitions of pairs of (Al,Ga)As layers of thickness $d$ with different concentrations of Al to vary their refractive index $n$. In the standard DBR, each layer fulfills the condition $nd = \lambda_{res}/4$, where $\lambda_{res}$ is the wavelength of the confined light. This condition allows unidirectional reflection of the light by optical interference. For the spacer layer, $n_{sp}d_{sp} = N\,\lambda_{res}/2$ where $N$ is an integer and $n_{sp}$ and $d_{sp}$ its refractive index and thickness, respectively. The MC is designed so that $\lambda_{res}$ matches the excitons emission wavelength.

The bottom semiconductor DBR consists of 25 repetitions of two alternating $Al_xGa_{1-x}As$ layers with nominal Al concentrations $x_1 = 0.15$ and $x_2 = 0.85$ and thicknesses $d_1 = 58.2$ nm and $d_2$ =66.6 m respectively. The different concentrations provide a nominal contrast in refractive index $\Delta n = n_1 - n_2$ V of around 13% at the resonance $\lambda_{res}$. The sample is designed and fabricated so that $\lambda_{res}$ is equal to that of the heavy-hole exciton (i.e. lowest energy) transition at 15 K (operating temperature). The bandwidth of the high reflectivity spectral region of the bottom DBR (i.e. the stopband) is around 70 nm at 15 K. The spacer layer is a $3\,\lambda/2$ thick layer containing three groups of 15 nm QWs located at the positions of maxima of the confined electromagnetic field. The measured transition energies of the heavy- and light-holes excitons in QWs at 15 K are 1.534 eV (808 nm) and 1.542 eV (804 nm), and their FWHM are 1.2 meV and 1.6 meV, respectively. The spacer layer is followed by 1.5 DBR pairs to protect the QWs from the highly energetic sputtering process.

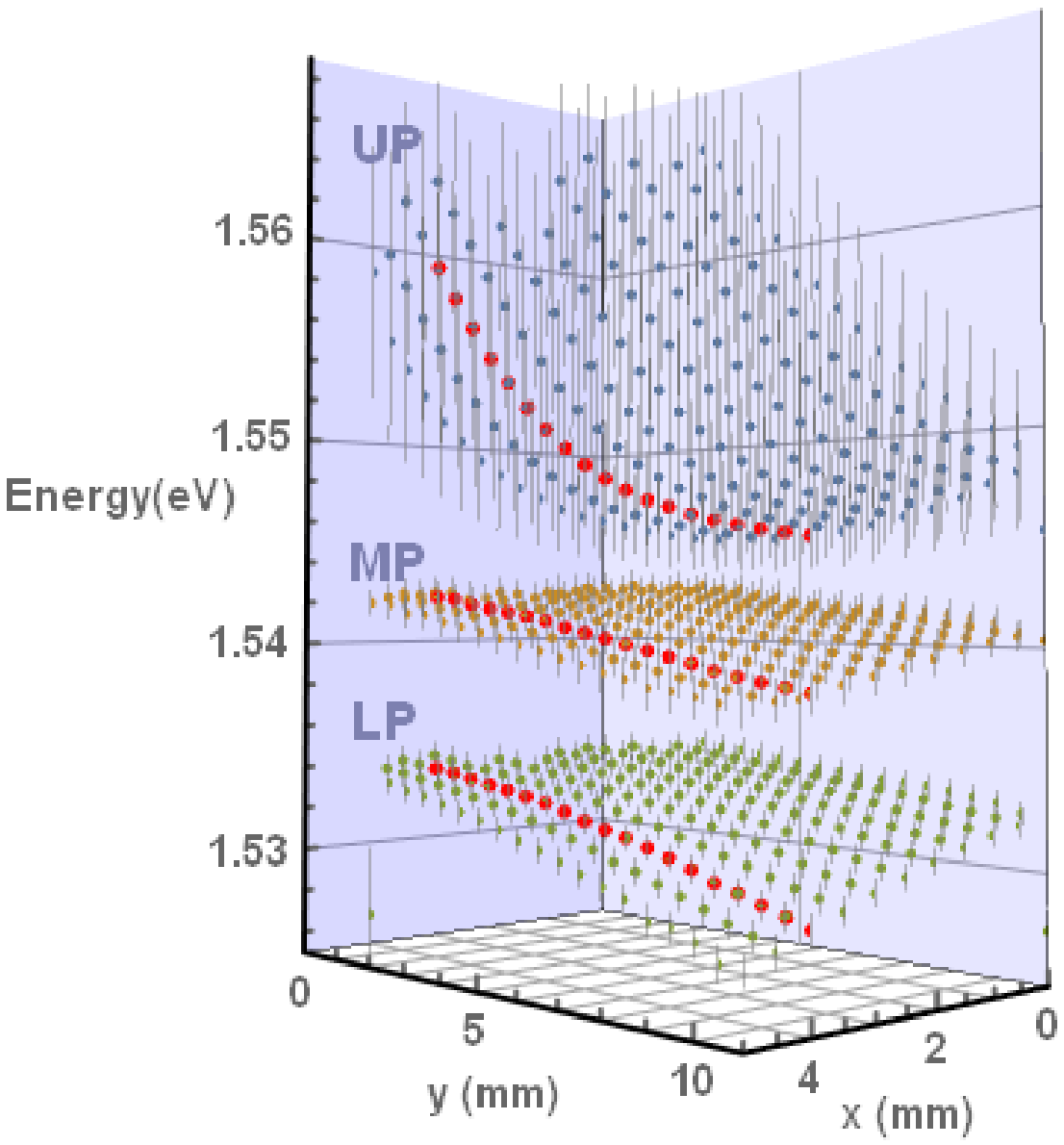

FIG A1. Analysis of the complete set of reflectivity measurements across the sample. The dots indicate the upper- (UP), middle- (MP) and lower-polariton (LP) energy values obtained from fitting the spectra with a three-Lorentzian oscillator model. The vertical gray bars indicate the corresponding FWHM extracted from the fit. The red dots indicate the data subset discussed throughout the manuscript.

The top DBR was deposited by rf-sputtering. The nominal refractive indexes of each material are 1.42 and 2.11, which yields a refractive index contrast $\Delta n = 36\%$. Thus, the number of layers to achieve large values of reflectivity $R$ is in principle much smaller than for the (Al,Ga)As-based semiconductor DBR. Calculations show that 6 pairs are sufficient to achieve high reflectivity. For the same reason, the bandwidth of the stopband is expected to be much larger (140 nm).

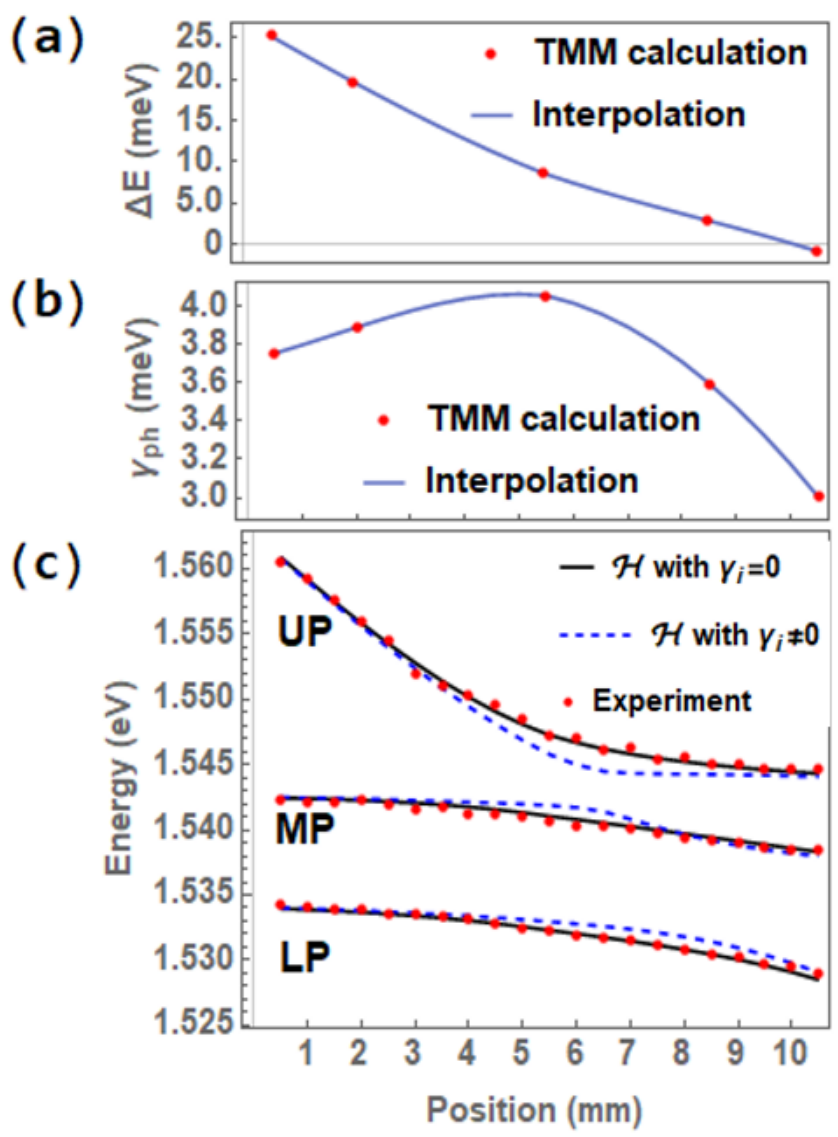

FIG C1. Position dependence of (a) the photon detuning $\Delta E$ and (b) its linewidth $\gamma_{ph}$ calculated using TMM (orange dots, see Fig. 2 in main text) and the numerical functions interpolated from them (blue lines). The functions are used in the simple three-level Hamiltonian $\mathcal{H}$ in eqn. 1. (c) Comparison of the experimental values (red dots) with the eigenvalues of $\mathcal{H}$ obtained with $\Omega_{hh} = 10\ meV$ and $\Omega_{lh} = 6.7\ meV$ with the linewidths terms $\gamma_i$ $(i = ph, hh, lh)$ set equal to 0 (black continuous line) and set to their experimental values.

The properties of the top dielectric DBR were determined by measuring the reflectance R over a broad wavelength range (600-900 nm) at room temperature. We made a measurement of the pure dielectric DBR at the border of the GaAs wafer, which is shielded from material deposition by the sample holder in the MBE reactor. By fitting this spectrum to TMM calculations, and with additional feedback from measurements on regions within the wafer, where the MC layers do exist, we estimated the real parameters of the sample. We determined that the optical thicknesses $nd$ of the SiO2 and Ta2O5 layers are, respectively, 2% and 14% smaller than expected, which translates into an overall blueshift of the stopband by 40 nm. Also, $\Delta n \sim 17\%$ (instead of the expected 36%), which yields a lower reflectivity. In large bandgap oxides such as the ones used, these values do not change significantly with temperature, so these parameters can be used for the simulation of the device response at the operating temperature of 15 K.

## VIII. APPENDIX C: THEORETICAL MODELS

In this Appendix we describe in detail the theoretical models used to analyze the experimental results discussed in Section III.

### C1. Three level interaction Hamiltonian

To diagonalize the three-level Hamiltonian $\mathcal{H}$ (eqn. 1 in main text) and estimate the interaction constants $\Omega_{hh}$ and $\Omega_{lh}$, we calculate numerical interpolation functions for the $y$-position dependent photon energy $E_{ph}(y)$ and linewidth $\gamma_{ph}(y)$ from the values extracted from the TMM calculation. These are shown in Figs. C1(a) and (b). The eigenvalues $\mathcal{H}$ are then fitted to the energies of the experimental reflectivity dips in Fig. 1 by adjusting $\Omega_{hh}$ and $\Omega_{lh}$, which are constant for all the detuning values. The results for $\Omega_{hh} = 10\ meV$ and $\Omega_{lh} = 6.7\ meV$ using this procedure are shown in Fig. C1(c). These values are consistent with what is typically reported for AlGaAs-based MCs. While the eigenvalues obtained from $\mathcal{H}$ with the imaginary components $\gamma_{hh}, \gamma_{lh}$ and $\gamma_{ph}$ set equal to 0 match the experimental curves, when they are set to their measured values, the fit deviates showing a smaller Rabi splitting for the same interaction constant. This is an expected effect as the imaginary parts are subtracted from the interaction constants reducing the observed Rabi splitting. Also, as discussed in the main text and shown in Fig. 3, the linewidths of the three branches are significantly overestimated.

To explore the origin of the anomalous linewidth behavior observed in our reflectivity experiments, we study the dependence of the polariton Hopfield coefficients for each branch, which can be calculated from the eigenvectors of $\mathcal{H}$. These are shown in Fig. C2. As discussed above, the model accurately reproduces the energy positions of the polariton branches, but fails to capture the observed linewidths, suggesting that additional physics is at play. Nevertheless, the Hopfield coefficients extracted from the fits provide insight into the internal composition of each polariton branch as a function of detuning. At large positive detuning, the upper branch is predominantly photonic, gradually mixing mostly with the light-hole exciton as the detuning decreases. The middle branch, initially dominated by the light-hole (LH) component, evolves into a mixed state involving all three constituents with nearly balanced contributions at $\Delta E \simeq 3.7\ meV$. The lower branch starts as a nearly pure heavy-hole exciton and increasingly hybridizes with the photon at smaller detunings. We suggest that this non-trivial mixing, particularly in the presence of two excitonic resonances, could lead to modified photon lifetimes and interference effects not captured in standard approaches, potentially explaining the observed linewidth reduction.

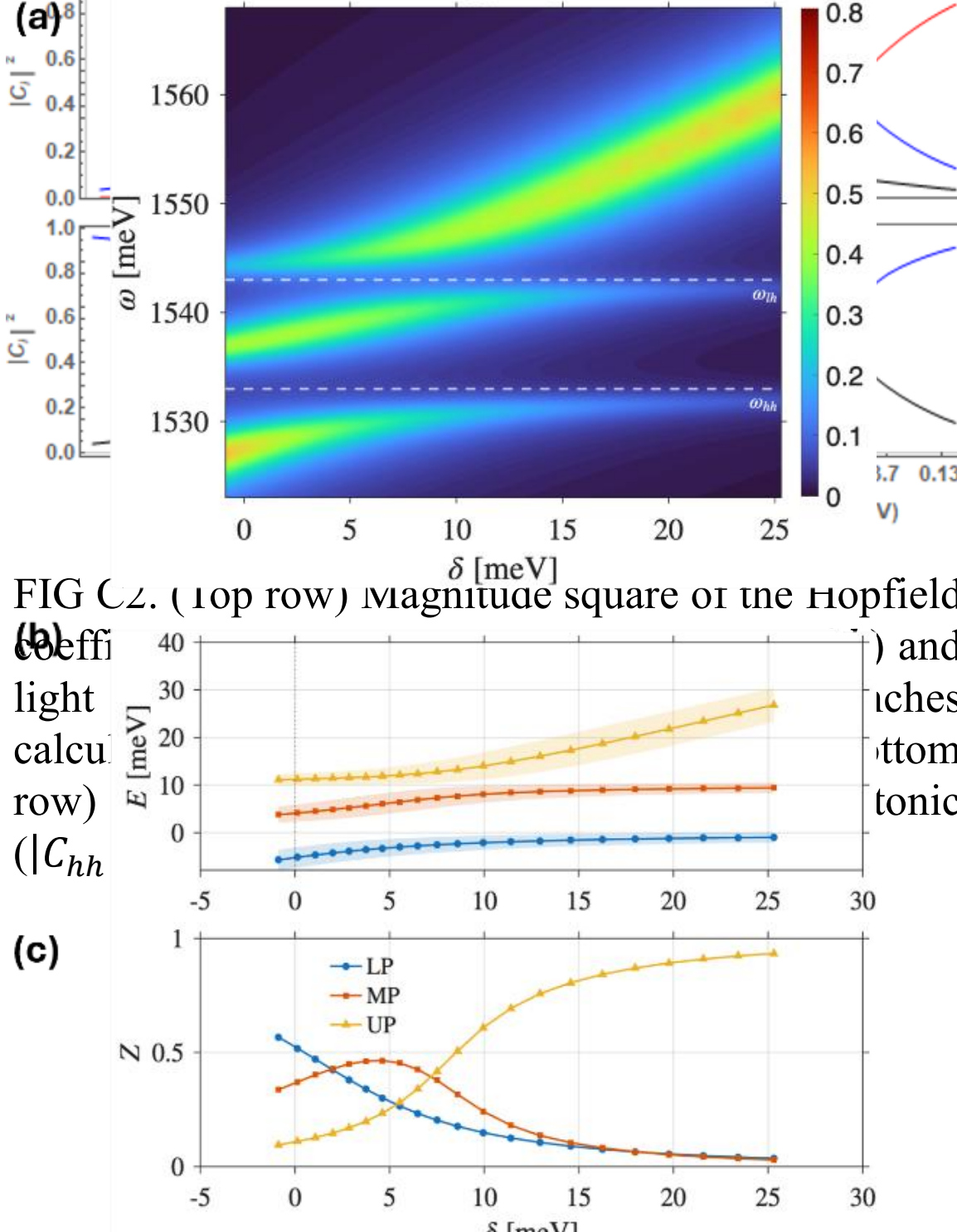

FIG C3. Photon spectral response and polariton parameters from Green-function poles. (a) Photon spectral function $\boldsymbol{A}cc(\delta,\omega)$. Three bright ridges reveal the lower (LP), middle (MP), and upper (UP) polariton branches as the cavity is tuned across the heavy- and light-hole excitons (horizontal dashed lines at $\omega hh$ and $\omega lh$). (b) Polariton energies $E_j(\delta)$ obtained from the complex poles of the cavity Green's function. The shaded bands indicate the branch half-widths $\gamma_j(\delta)$. (c) Photon-channel residues $Z_j(\delta)$ (Hopfield photon weights), given by the residues of$G_{cc}$ at the poles.

**C2. Retarded Green's function formalism (Dyson equation)**

The starting point is the retarded cavity Green's function

$$G_{cc}(\omega) = \frac{1}{\omega - E_{ph}(\delta) + i\gamma_{ph} - \Sigma(\omega)}$$

where $E_{ph}(\delta)$ and $\gamma_{ph}$ denote the bare cavity resonance and damping, respectively. The photon self-energy $\Sigma(\omega)$ accounts for the coupling to the heavy- and light-hole excitons,

$$\Sigma(\omega) = \frac{|\Omega_{hh}|^2}{\omega - E_{hh} + i\gamma_{hh}} + \frac{|\Omega_{lh}|^2}{\omega - E_{lh} + i\gamma_{lh}},$$

with $\Omega_i$ the Rabi couplings and $\gamma_i$ the excitonic linewidths. The quasiparticle approach consists into summarizing the main properties into few renormalized quantities. Here, we study the quasiparticle energy $E_j$, residue $Z_j$ and damping rate $\Gamma_j$. The idea of this approach is that around the quasiparticle poles, the Green's function is given by

$$G_{cc}(\omega) \simeq \frac{Z_j}{\omega - E_j + i\Gamma_j}.$$

In the presence of dissipation, the poles lie in the complex plane $\omega = E_j - i\Gamma_j$ where $E_j$ is the quasiparticle energy and $\Gamma_j$ the damping rate. On the other hand, the quasiparticle residue $Z_i$ accounts for the spectral weight of each quasiparticle branch. The quasiparticle residue is given by

$$Z_j = [1 - \partial_\omega \mathrm{Re}\,\Sigma(\omega)]^{-1}_{\omega=E_j}$$

The photon-weight average damping rate is given by

$$\Gamma_T = \sum_{i=1}^{3} Z_i \Gamma_i,$$

which in general can strongly differ from the bare photon damping rate $\gamma_c$ and the sum of the damping rates $\sum_j \Gamma_j = \gamma_c + \gamma_{hh} + \gamma_{lh}$. Indeed, in principle, the total photonic damping, may vary significantly with detuning, since the contributions from the polariton modes are dynamically renormalized rather than algebraically redistributed. This explains why the photon-like upper polariton can undergo pronounced narrowing without a corresponding broadening of the other branches.

Figure C3 summarizes our analysis as a function of the cavity detuning $\delta$. Panel (a) displays the photon spectral function $\mathcal{A}_{cc}(\delta,\omega)$. Three bright lines identify the lower (LP), middle (MP), and upper (UP) polariton branches that emerge as the cavity is tuned across the heavy- and light-hole excitons (horizontal dashed lines at $\omega_{hh}$ and $\omega_{lh}$). Because $\mathcal{A}_{cc}$ is projected onto the cavity mode, the intensity directly reflects the photon content of each branch: more photonic branches appear brighter, and the avoided crossings with each exciton are clearly visible.

Panel (b) shows the dispersions $E_j(\delta)(j = LP, MP, UP)$ obtained from the complex poles of the photon Green's function, $G_{cc}(\omega)$. Specifically, the pole positions $\omega_j(\delta) = E_j(\delta) - i\gamma_j(\delta)$ yield both the branch energies; the shaded envelopes indicate the damping rate of the polariton branches. The three-branch dispersion reproduces the expected avoided crossings; as detuning increases, the UP acquires a stronger photonic character and shifts upward, the LP bends toward the lower exciton, and the MP remains in between with a comparatively moderate dispersion.

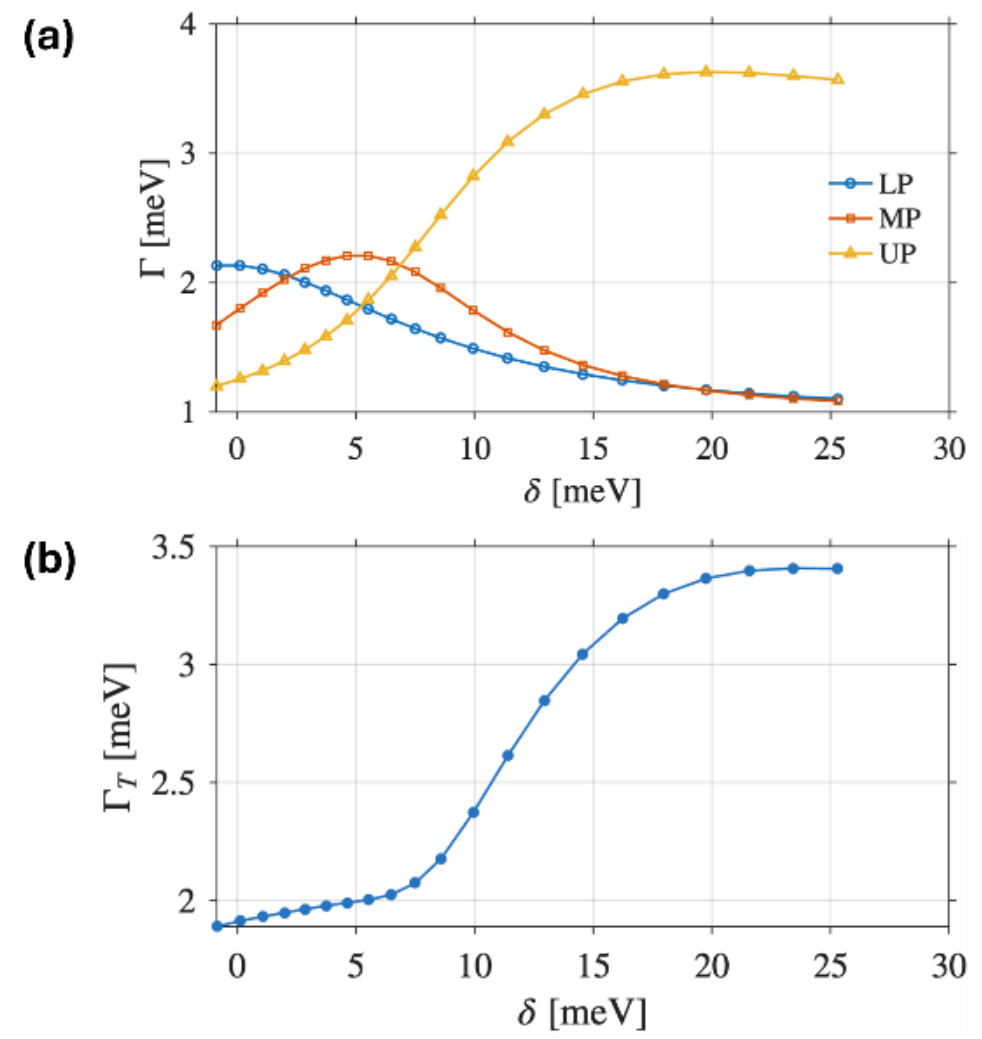

FIG C4. Polariton damping and photon-weighted total damping. (a) Branch damping rates $\Gamma_j(\delta)$ $(j = LP, MP, UP)$. ( b) Effective photon damping $\Gamma_T(\delta)$.

Panel (c) illustrates the photon-channel quasiparticle residues $Z_j(\delta)$ (Hopfield photon weights), obtained from the residues of $G_{cc}$ at the poles. A smooth transfer of photon weight from LP to UP with increasing $\delta$ is evident, while the MP exhibits a dome-like photon content peaking near intermediate detuning values. The residues satisfy $\sum_j Z_j \simeq 1$ and control the intensity in $A_{cc}$.

Finally, in Fig. C4 we show the polariton damping analysis versus detuning $\delta$. Panel (a) shows the damping rates $\Gamma_j(\delta)$ for the lower (LP), middle (MP), and upper (UP) polaritons, obtained directly from the complex poles of the cavity Green's function $G_{cc}(\omega)$. With increasing detuning $\Gamma_{LP}$ decreases, $\Gamma_{UP}$ grows markedly, and $\Gamma_{MP}$ exhibits a dome-like behaviour peaking near the exciton–cavity resonance, reflecting the redistribution of photonic/excitonic character among the branches. Panel (b) shows a one-number summary for the cavity photons, the residue-weighted total damping $\Gamma_T(\delta)$. The monotonic rise of $\Gamma_T$ tracks the transfer of photon weight from LP to UP and the associated redistribution of decay channels.

## ACKNOWLEDGMENTS

E.A.C-M. would like to acknowledge the financial support of DAAD through project 91807398 of the Re-invitation Program for Former Scholarship Holders, 2023; SECIHTI (formerly CONAHCYT), through project FC-2016-2183 and IICO-UASLP. A.C-G. acknowledges financial support from UNAM DGAPA PAPIIT Grant No. IA101325, Project CONAHCYT No. CBF2023-2024-1765 and PIIF25. E.A. C-M. and Y.G.R. acknowledge funding support from DGAPA-UNAM, Grant PAPIIT No. IN108524.

[1] C. Weisbuch, M. Nishioka, A. Ishikawa, and Y. Arakawa, Observation of the coupled exciton-photon mode splitting in a semiconductor quantum microcavity, Phys. Rev. Lett. **69**, 3314 (1992).

[2] H. Deng, G. Weihs, C. Santori, J. Bloch, and Y. Yamamoto, Condensation of Semiconductor Microcavity Exciton Polaritons, Science **298**, 199 (2002).

[3] J. Kasprzak et al., Bose-Einstein condensation of exciton polaritons, Nature **443**, 409 (2006).

[4] R. Balili, V. Hartwell, D. Snoke, L. Pfeiffer, and K. West, Bose-Einstein Condensation of Microcavity Polaritons in a Trap, Science **316**, 1007 (2007).

[5] T. Byrnes, N. Y. Kim, and Y. Yamamoto, Exciton–polariton condensates, Nature Physics **10**, 803 (2014).

[6] I. Carusotto and C. Ciuti, Quantum fluids of light, Reviews of Modern Physics (2013).

[7] F. I. Moxley, E. O. Ilo-Okeke, S. Mudaliar, and T. Byrnes, Quantum technology applications of exciton-polariton condensates, Emergent Mater. **4**, 971 (2021).

[8] A. Kavokin, T. C. H. Liew, C. Schneider, P. G. Lagoudakis, S. Klembt, and S. Hoefling,

Polariton condensates for classical and quantum computing, Nat Rev Phys **4**, 7 (2022).

[9] D. G. Angelakis, editor , *Quantum Simulations with Photons and Polaritons* (Springer International Publishing, Cham, 2017).

[10] T. Boulier et al., Microcavity Polaritons for Quantum Simulation, Advanced Quantum Technologies **3**, 2000052 (2020).

[11] F. Toffoletti and E. Collini, Coherent phenomena in exciton–polariton systems, J. Phys. Mater. **8**, 022002 (2025).

[12] D. Sanvitto and S. Kéna-Cohen, The road towards polaritonic devices, Nature Materials **15**, 1061 (2016).

[13] V. Ardizzone, M. L. De, G. M. De, L. Dominici, D. Ballarini, and D. Sanvitto, Emerging 2D materials for room-temperature polaritonics, Nanophotonics **8**, 1547 (2019).

[14] H. Kang, J. Ma, J. Li, X. Zhang, and X. Liu, Exciton Polaritons in Emergent Two-Dimensional Semiconductors, ACS Nano **17**, 24449 (2023).

[15] B. Xiang and W. Xiong, Molecular Polaritons for Chemistry, Photonics and Quantum Technologies, Chem. Rev. **124**, 2512 (2024).

[16] R. Su, A. Fieramosca, Q. Zhang, H. S. Nguyen, E. Deleporte, Z. Chen, D. Sanvitto, T. C. H. Liew, and Q. Xiong, Perovskite semiconductors for room-temperature exciton-polaritonics, Nat. Mater. **20**, 1315 (2021).

[17] C. P. Dietrich, A. Steude, L. Tropf, M. Schubert, N. M. Kronenberg, K. Ostermann, S. Höfling, and M. C. Gather, An exciton-polariton laser based on biologically produced fluorescent protein, Science Advances **2**, e1600666 (2016).

[18] M. Nakayama, M. Kameda, T. Kawase, and D. Kim, Cavity polaritons of heavy-hole and light-hole excitons in a CuI microcavity, Phys. Rev. B **83**, 235325 (2011).

[19] D. C. T. Ferreira, A. C. S. P. Pimenta, and F. M. Matinaga, Light and heavy hole exciton polariton Faraday rotation in a single GaAs microcavity, J. Phys.: Conf. Ser. **864**, 012082 (2017).

[20] H. Alnatah, S. Liang, Q. Wan, J. Beaumariage, K. West, K. Baldwin, L. N. Pfeiffer, M. C. A. Tam, Z. R. Wasilewski, and D. W. Snoke, *Strong Coupling of Polaritons at Room Temperature in a GaAs/AlGaAs Structure*, arXiv:2502.12338.

[21] H. Alnatah, S. Liang, Q. Yao, Q. Wan, J. Beaumariage, K. West, K. Baldwin, L. N. Pfeiffer, and D. W. Snoke, Bose–Einstein Condensation of Polaritons at Room Temperature in a GaAs/AlGaAs Structure, ACS Photonics **12**, 48 (2025).

[22] Y. A. García Jomaso, B. Vargas, D. L. Domínguez, R. J. Armenta-Rico, H. E. Sauceda, C. L. Ordoñez-Romero, H. A. Lara-García, A. Camacho-Guardian, and G. Pirruccio, Intercavity polariton slows down dynamics in strongly coupled cavities, Nat Commun **15**, 2915 (2024).

[23] Atac Imamoglu, R. J. Ram, S. Pau, and Y. Yamamoto, Nonequilibrium condensates and lasers without inversion: Exciton-polariton lasers, Phys. Rev. A **53**, 4250 (1996).

[24] M. Steger, B. Fluegel, K. Alberi, D. W. Snoke, L. N. Pfeiffer, K. West, and A. Mascarenhas, Ultra-low threshold polariton condensation, Opt. Lett., OL **42**, 1165 (2017).

[25] D. M. Whittaker, P. Kinsler, T. A. Fisher, M. S. Skolnick, A. Armitage, A. M. Afshar, M. D. Sturge, and J. S. Roberts, Motional Narrowing in Semiconductor Microcavities, Physical Review Letters **77**, (1996).

[26] A. V. Kavokin, Motional narrowing of inhomogeneously broadened excitons in a semiconductor microcavity: Semiclassical treatment, Phys. Rev. B **57**, 3757 (1998).

[27] V. Savona, C. Piermarocchi, A. Quattropani, F. Tassone, and P. Schwendimann, Microscopic Theory of Motional Narrowing of Microcavity Polaritons in a Disordered Potential, Phys. Rev. Lett. **78**, 4470 (1997).

[28] L. C. Flatten, Z. He, D. M. Coles, A. A. P. Trichet, A. W. Powell, R. A. Taylor, and J. M. Smith, Room-temperature exciton–polariton coupling in a monolayer of $MoSe_2$, Nature Communications **7**, 13734 (2016).

[29] S. Kéna-Cohen and S. R. Forrest, Room-temperature polariton lasing in an organic single-crystal microcavity, Nature Photonics **4**, 371 (2010).

[30] T. C. H. Liew, R. Su, S. Gong, X. Liu, and T. C. Sum, Polariton devices: hybrid perovskites enter the arena, Nature Materials **20**, 1125 (2021).

[31] J.-G. Rousset et al., Strong coupling and polariton lasing in Te based microcavities embedding (Cd,Zn)Te quantum wells, Applied Physics Letters **107**, 201109 (2015).

[32] D. C. Teles, A. C. S. Pimenta, and F. M. Matinaga, *Light and Heavy Hole Exciton Polariton Faraday Rotation in a Single GaAs Microcavity*, in *Photonics and Fiber Technology*

*2016 (ACOFT, BGPP, NP) (2016), Paper JT4A.23* (Optica Publishing Group, 2016), p. JT4A.23.

[33] K. Santhosh, O. Bitton, L. Chuntonov, and G. Haran, Vacuum Rabi splitting in a plasmonic cavity at the single quantum emitter limit, Nat. Commun. **7**, 11823 (2016).

[34] R. Chikkaraddy, B. de Nijs, F. Benz, S. J. Barrow, O. A. Scherman, E. Rosta, A. Demetriadou, P. Fox, O. Hess, and J. J. Baumberg, Single-molecule strong coupling at room temperature in plasmonic nanocavities, Nature **535**, 127 (2016).

[35] J. Qin, Y.-H. Chen, Z. Zhang, Y. Zhang, R. J. Blaikie, B. Ding, and M. Qiu, Revealing Strong Plasmon-Exciton Coupling between Nanogap Resonators and Two-Dimensional Semiconductors at Ambient Conditions, Phys. Rev. Lett. **124**, 063902 (2020).

[36] D. G. Baranov, M. Wersäll, J. Cuadra, T. J. Antosiewicz, and T. Shegai, Ultrastrong coupling between nanoparticle plasmons and cavity photons at ambient conditions, Nat. Commun. **11**, 2715 (2020).

[37] T. Wu et al., Ultrastrong exciton-plasmon couplings in WS2 multilayers synthesized with a random multi-singular metasurface at room temperature, Nat Commun **15**, 3295 (2024).

[38] C. C. Katsidis and D. I. Siapkas, General transfer-matrix method for optical multilayer systems with coherent, partially coherent, and incoherent interference, Appl. Opt. **41**, 3978 (2002).

[39] P. Kinsler and D. M. Whittaker, Linewidth narrowing of polaritons, Phys. Rev. B **54**, 4988 (1996).

[40] V. Kravtsov et al., Nonlinear polaritons in a monolayer semiconductor coupled to optical bound states in the continuum, Light Sci Appl **9**, 56 (2020).